# Generating Nanoporous Graphene from Point and Stone-Wales Defects: A Study with Dimensionally Restricted Molecular Dynamics (DR-MD)

Ji Wei Yoon[1,2*]

[1]Institute for Infocomm Research (I$^2$R), Agency for Science, Technology and Research (A*STAR), 1 Fusionopolis Way, #21-01, Connexis South Tower, Singapore 138632, Republic of Singapore

[2]Department of Physics, Imperial College London, South Kensington Campus, London SW72AZ, United Kingdom

*Corresponding author

Email: yoon_ji_wei@i2r.a-star.edu.sg

## Abstract

Defects in graphene are both a boon and a bane for applications – they can induce uncontrollable effects but can also provide novel ways to manipulate the properties of pristine graphene. Nanoporous Graphene, which contains nanoscopic holes, has found impactful applications in sustainability domains, e.g. gas separation, water filtration membranes and battery technologies. For this report, we investigate pore formation in graphene with no defect, one and two mono-vacancies, and two di-vacancies using bespoke Dimensionally Restricted Molecular Dynamics (DR-MD) designed for the purpose. We show DR-MD to be superior to free-standing or substrate suspended configurations for simulating stable defected structures. Applying DR-MD, stable pore configurations are identified, and their formation mechanisms proposed. We also investigated formation mechanisms due to two Stone-Wales 55-77 defects, and the formation energies of their linearly extended structures, along the zigzag and armchair directions, and when they are placed in different relative orientations. This study offers a way to identify stable porous defect structures in graphene and insights into atomistic pore formation mechanisms for an environmentally important material.

## Introduction

In recent years there has been an explosion of research activities on graphene and other related carbon systems [1–4]. This is due, in large part, to the search for new materials for microelectronic applications, in expectation of the coming post-silicon era. Graphene possesses many outstanding electronic [5,6], thermal [6], and mechanical [7] properties which make it a good candidate for such purposes. These properties also make it useful for many novel applications, like storage medium for hydrogen [8–10], DNA detector [11,12] for disease detection and filtration membranes for low-cost water treatment [13], gas separation [14] and battery applications [15], to state a few.

In membrane applications, Graphene-based porous materials are found to be better than other porous carbon materials for a few reasons: (1) High mechanical strength that confers structural stability, (2) Resistance to chemical and thermal degradation in harsh environments, (3) High conductivity that makes it an ideal charge collector across the layer. It is known that the properties of graphene are modified significantly by the presence of structural defects, amongst which point defects are ones of the most common. In this paper, two common types of defects that will lead to pore formation – Stone-Wales defects and vacancies – are investigated.

A Stone-Wales defect is generated by a rearrangement of lattice atoms in graphene, without a gain or loss of atoms. Since it involves formation pathways with energies usually an order of magnitude smaller as compared to that of vacancies, it is relatively more prevalent. Since Stone-Wales defects involve rings with larger number of members than the pristine hexagon, they are a source of pores in graphene. Also, the possibility of having such defects means that one may be able to control the properties of graphene through defect engineering. Therefore, there is a need to identify and characterize them.

A vacancy is a missing lattice atom. A lattice atom can be deliberately removed by chemical reaction or particle irradiation. When two neighboring atoms on the lattice are missing a di-vacancy defect is formed. Since defects are known to migrate and interact, there is a need to understand the details of their interaction and the various structures, e.g. pores, that can be borne from such interactions. This information will inform further experimental effort in defect identification, characterization and engineering.

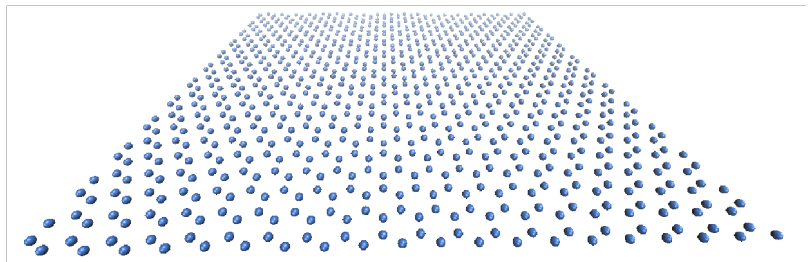

**Figure 1** An energy minimized graphene sheet with 960 atoms. Note that the x direction is from left to right (i.e. along the zigzag direction of the sheet), y direction is along the perpendicular armchair direction, and the z direction is along the out-of-plane direction. This sheet of graphene is the source configuration from which other defected graphene sheet used in the other simulations were generated.

There have been extensive simulations conducted on idealized porous Graphene-based membranes to elucidate the impact of pore size, distribution and porosity on filtration characteristics[16–20]. However, these studies typically investigate a particular porous configuration and do not account for dynamical introduction of more nanopores into the membranes during actual applications. This is due to the complexity involved for the required simulations and the short time-scales typically accessible by Molecular Dynamics (MD) simulations. Applying DR-MD, we show in this work that there are many stable porous structures consistent with applicative conditions. Therefore, there is a need to anticipate their formation and nanostructure, which will inform ways to promote/demote their occurrence and the associated performance shifts and degradation in in situ environments.

In view of the mentioned gaps in previous studies of nanoporous Graphene, this report is focused on three main questions:

1. What are the optimal computational parameters and simulation conditions to generate energetically stable porous defect structures?
2. What are the stable porous defect structures and their formation mechanisms of the graphene systems of interest (to be elaborated below)?
3. What is the impact of the 55-77 Stone-Wales defect structures on the existence of nanopores in Graphene?

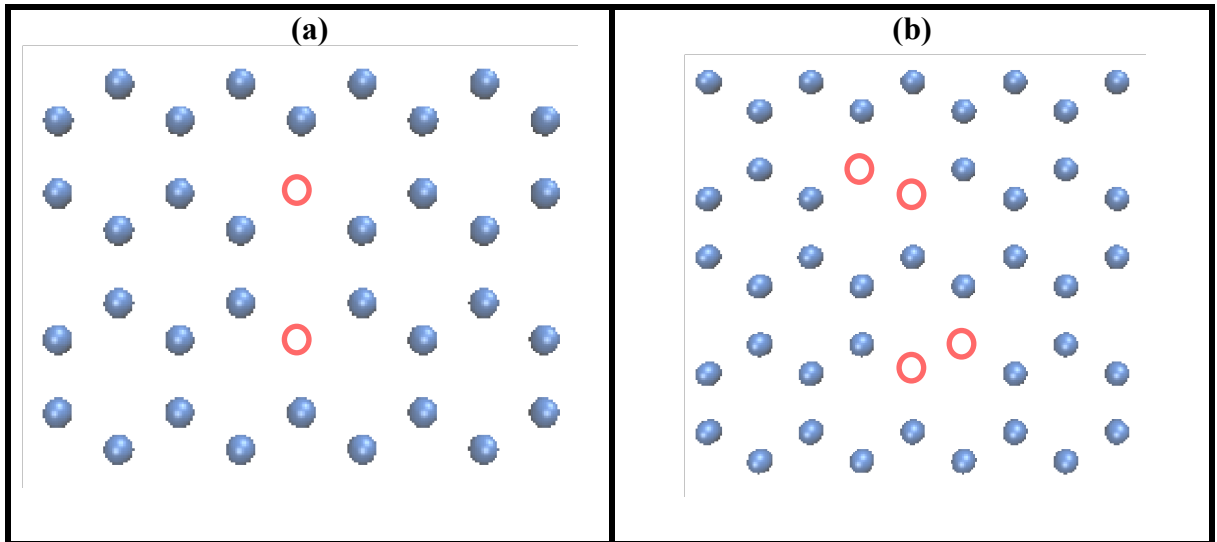

**Figure 2** Schematics of defected graphene sheets. (a) A blown-up defected area where there are two mono-vacancies which are separated by two lattice sites. Pink circles indicate the locations of the two mono-vacancies. (b) A blown-up defected area where there are two di-vacancies which are separated by two lattice sites. Pink circles indicate the locations of the two di-vacancies.

The first question was investigated through simulations of a sheet of graphene with two divacancies placed at different orientations and separation distances. FIG. 2(b) shows one of the possible configurations of two di-vacancies. Independent sets of simulations were carried out on these different configurations to investigate the difference in bond dynamics between the graphene sheets that were left (A) free-standing, (B) only allowed to have in-plane movements (z-restricted) and (C) suspended on a sheet of rigid pristine graphene. The temperature ramp method was employed to drive the bond dynamics. The goal was to identify the best amongst the three methods for generating stable structures and to discover novel structures formed from the coalescence of two di-vacancies.

Then, the best method was employed to answer the second question for the cases of (1) pristine graphene (refer to FIG. 1 for an illustration), (2) graphene with a single mono-vacancy and those with (3) two mono-vacancies (refer to FIG 2(a) for an illustration).

From our analysis of these simulation results, the Stone-Wales 55-77 defect was deemed to be one of the more prevalent defects. Hence, further simulations were performed to investigate the energetics of linear structures formed from multiple 55-77 defects. The energetics of the interaction between two 55-77 defects was also investigated. For these simulations, a distinction was made between defect structures located along the armchair (AC) and zigzag (ZZ) directions of the parent graphene.

## Methodologies

For our MD simulations, ReaxFF force field [21,22] was employed as implemented in LAMMPS [23] by Sandia National Laboratory.

The ReaxFF force field has been chosen because it allows for the formation and breakage of bonds in chemical reactions. It is a bond-order potential that considers the interplay between coordination, bond strength and bond length, enabling evolution between single, double and conjugated bonds. Moreover, it has been used extensively [24–26] for hydrocarbon systems and yielded results that compare well with experimental values.

LAMMPS is employed as it is specifically written to run in parallel on multiple cores, enabling massive speed-up over serial computation. This allowed the investigation of a system of 960 carbon atoms with a runtime of approximately an hour for a simulation time of 1 picosecond on an Intel i7 Quad core machine. Such a length of simulation time was found to be sufficient for most of the bond dynamics of interest.

Visual Molecular Dynamics (VMD) [27] is a molecular graphics software which also has the inbuilt capability for generating pristine graphene sheets of any user-specified sheet dimensions. The decision has been made to use a sheet of square pristine graphene as the basis from which all other defected initial configurations for the simulations were generated. It was chosen to have 960 carbon atoms so as to possess a large area where extended defects structures could be investigated without the problem of self-interaction across the periodic boundaries. VMD has been used to generate the graphene sheet and to visualize the output of the simulations.

In all of the simulations, the simulation cell was split into two boxes, each with approximately 480 carbon atoms to run on a single core. This was deemed to be the optimal setting which allowed the most significant computational speed-up. Further splitting of the simulation box increased the time needed for data communication, so increasing the runtime. The NVE integration was used to create system trajectory consistent with the microcanonical ensemble.

To ascertain the optimal time step, a sheet of graphene with a single mono-vacancy was simulated using a time step of 0.1, 1.0 and 10.0 fs with a temperature ramp from 0K to 300K over 1 ps. The Berendsen thermostat [28] was used in all the simulations for temperature control.

With the choice of 10 fs time step, the graphene sheet was observed to disintegrate beyond a critical temperature. This can be explained as follows: The thermalization process involves giving the ensemble of carbon atoms a distribution of velocities. The velocities of the atoms are then integrated over the time step to yield their displacements. As the temperature is increased, there exist an increased number of carbon atoms within the ensemble which have velocities that integrated to displacements larger than the average carbon-carbon bond length. If such greatly displaced carbon atoms occur at low densities, the graphene sheet retains its ability to recover its periodicity by a rearrangement of the lattice atoms. However, the sheet of graphene loses this ability at a critical temperature where there exists too high a density of greatly displaced atoms. This results in the disintegration of the sheet. Note that this effect, as will be shown in the simulations with shorter time steps, is purely a stability issue of the algorithm at large time steps and has no physical significance.

With the choice of 1.0 fs time step, no bond dynamics was observed. When 0.1 fs time step was used, a formation of bonds around the mono-vacancy site and a reduction in the number of dangling bonds were seen. Effectively, a structure of lower energy was formed. The less energetic defect structure formed and the increased bond dynamics suggest that 0.1 fs is a more suitable time step for further simulations.

In order to prevent edge effects, periodic boundaries were used in the x-y plane. It is important to remove edges in the simulations because edge effects have been known to have significant effect on the energetics and the conformation [29] of 2D graphene systems. These additional warping effects from edges would have an impact on the investigation of defect structures, whose interaction with the sheet itself depends on its alteration of the local morphology. The determination of the optimal size of the simulation cell was deemed important because only with an approximately stress and strain-free system could one expect to see dynamics that are truly reflective of free-standing graphene. To further prevent self-interaction across the periodic boundaries, systems with low number of defects were simulated. This implies that the stress and strain produced by a defect is expected to be limited to the vicinity of the defect. Therefore, this allowed the use of a static periodic boundary in the two planar directions. The length and width of the simulation cell was determined from the dimensions of an energy minimized pristine graphene that was kept planar – the idea is that any change in morphology of the pristine sheet at finite temperature - which must occur for a 2D crystal to remain stable and avoid disintegration as predicted by the Mermin-Wagner [30] theorem - will be like standing waves on the surface of the skin of a drum – fluctuations upon the planar configuration. Thus, the cell size was determined to be 50.0 Å in the x direction and 51.9 Å in the y direction. The sheet of graphene was placed on the x-y plane and the walls of the simulation box in the z direction were given a fixed non-periodic boundary condition.

The Polak-Ribiere version of the conjugate gradient (CG) algorithm has been employed for all energy minimization procedures. The energy tolerance was set to be 1 part in $10^{10}$. This accuracy was chosen as a compromise against runtime. An energy minimization run for 960 carbon atoms took around 1.5 hours on 2 Intel i7 CPU cores.

**Results and Discussion**

Within each subsection of this section, the setup of the molecular dynamics simulations will be discussed alongside with the results. A complete discussion is done in each subsection because the results of subsection (I) directly inform the setup of the simulations discussed in the next two subsections. Additionally, the investigation carried out in subsection (IV) was informed by the results of (II) and (III).

I.  **Double Di-vacancies**

Sheets of graphene with different arrangements of two di-vacancies were generated by removing carbon atoms from an energy minimized pristine graphene (FIG. 1.). FIG. 2(b) illustrates one such arrangement. Each of these defected graphene sheets was simulated under three different conditions. The three sets of simulations involve the graphene being free-standing, z-restricted (or, having no out-of-plane movement of atoms) and suspended on a rigid planar sheet of pristine graphene, respectively. The purpose of all simulations was to identify stable defect structures, then to use that information to identify the best simulation conditions for the purpose.

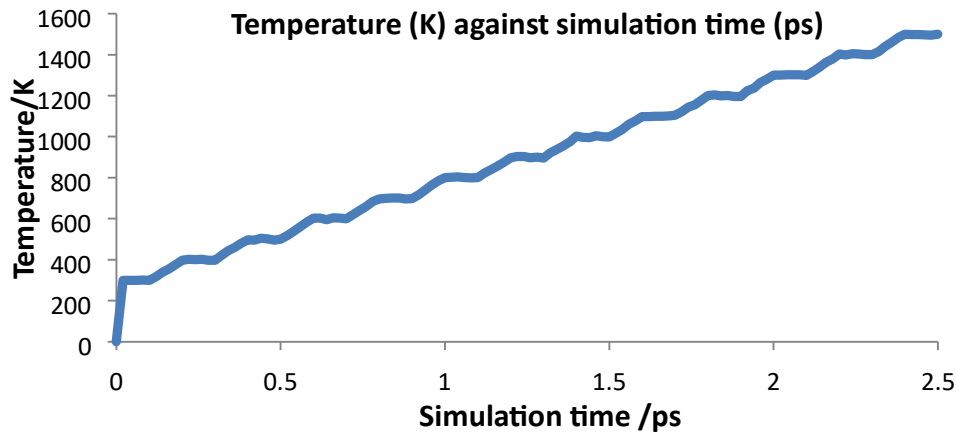

**Figure 3** The graph of the temperature plotted against simulation time of a typical defected graphene sheet during temperature ramp. Note that the linear temperature ramp was disrupted at a regular interval of 0.1 ps, where the temperature was kept constant for 0.1 ps for the graphene sheet to react to the higher temperature. Only the first 2.5 ps of the simulation was plotted to reveal the fine details of the ramp.

Free-standing graphene sheets were considered in the first set of simulations because of its relevance to many experiments [31–33]. Moreover, more elaborate bond dynamics was expected due to the presence of an additional degree of freedom. It was hypothesized that enhanced bond dynamics would improve the atoms' ability to rearrange themselves into a stable configuration.

A planar rigid sheet of pristine graphene was used as a substrate for the defected graphene in the second set of simulations. This system is simple but the results can be useful to aid our understanding of the widely studied graphitic systems.

The limiting case of substrate suspended graphene is z-restricted graphene, whose atoms can be thought of as experiencing an unbreakable and isotropic attraction to a substrate. The third set of simulations was performed with such a condition. It was hypothesized that such restriction will result in greater in-plane motion of the atoms, leading to greater bond dynamics and more stable configurations.

In all of the simulations, the graphene sheet was subjected to a temperature ramp from 300K to 4000K in 7.5 ps. FIG. 3 is a plot of the temperature of a typical defected graphene sheet against simulation time during the temperature ramp. Intervals of constant temperature were periodically introduced along the ramp to give the system enough time to respond and allow one to identify any novel defect structure made accessible by the higher temperature.

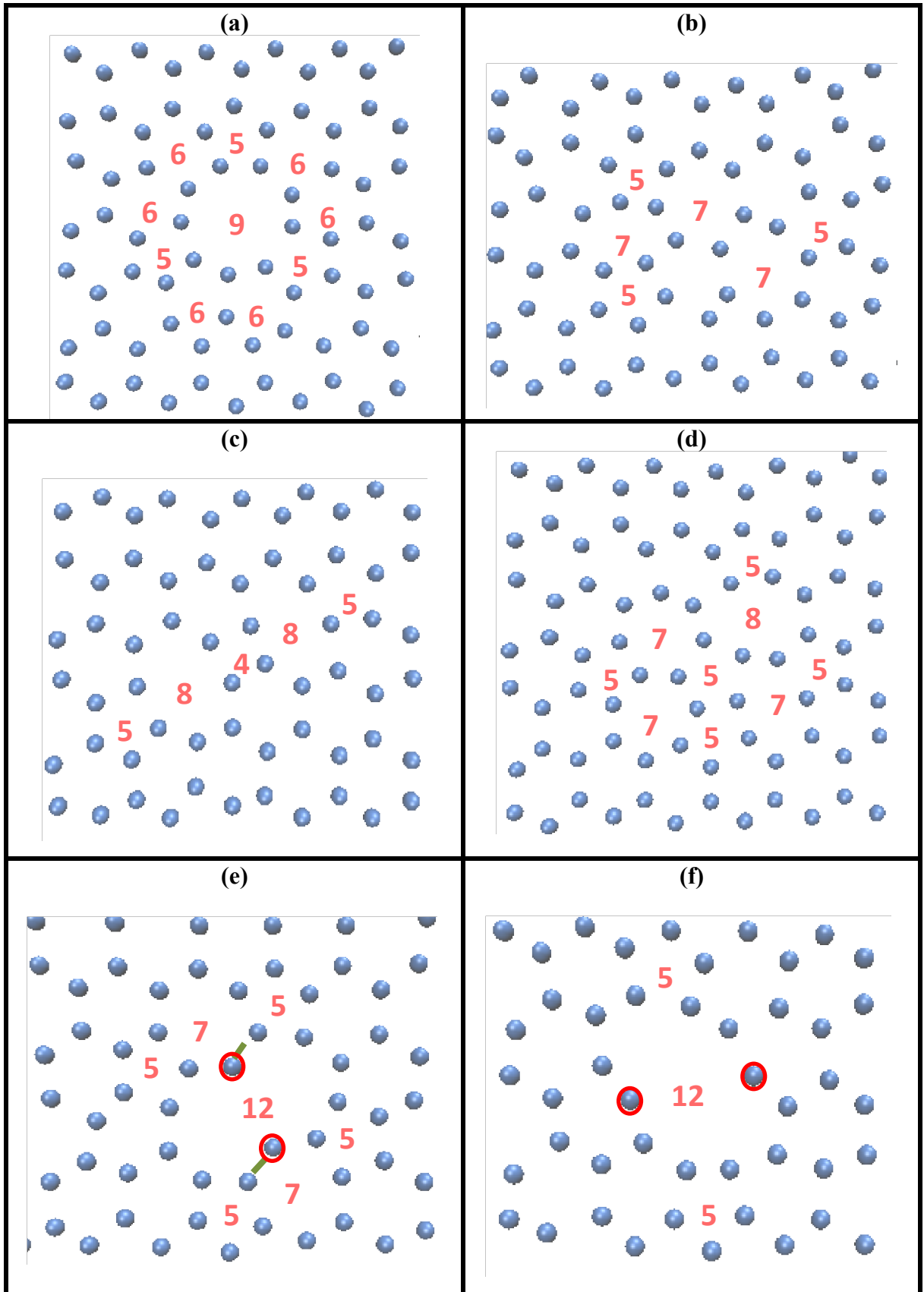

**Figure 4** Defected Graphene structures with double di-vacancies. (a) A 555-9-(66)₃ defect formed from two divacancies. This was observed at ~4000K in z-restricted graphene and ~2500K in free-standing graphene. The same structure was observed in a more sophisticated

DFT simulation [34]. (b) A defect formed from two di-vacancies. It has only 5 and 7 membered rings. This was observed in z-restricted graphene at ~4000K. (c) A 5-8-4-8-5 defect formed from two divacancies. This was observed in z-restricted graphene at ~4000K. The same structure has been observed experimentally [35]. (d) A defect observed in z-restricted graphene at ~3700K. It was formed from two di-vacancies. (e) A defect formed from two di-vacancies. It has a 12 membered pore surrounded by four hexagons and two 5-7-5 defects. Atoms with dangling bonds are indicated by red circles. This was observed in substrate-suspended graphene at ~1800K and freestanding graphene at ~2700K. (f) A 5-12-5 defect formed from two divacancies. This was observed in z-restricted graphene at ~4000K. This structure can be formed from that of FIG. 4(e) by rotating the two bonds indicated by the green dotted lines. The red circles indicate atoms with dangling bonds.

FIG. 4(a), 4(b), 4(c) and 4(d) illustrate all the observed configurations with no dangling bonds. The absence of dangling bonds suggests that these structures are energetically stable and thus should be observed in experiments.

FIG. 4(e) and 4(f) illustrate all the configurations with two dangling bonds and approximate rotational symmetry. These structures are relatively stable than those with more dangling bonds and so might be observed in experiments. Indeed, defect structures with dangling bonds have been observed in experiments [36,37].

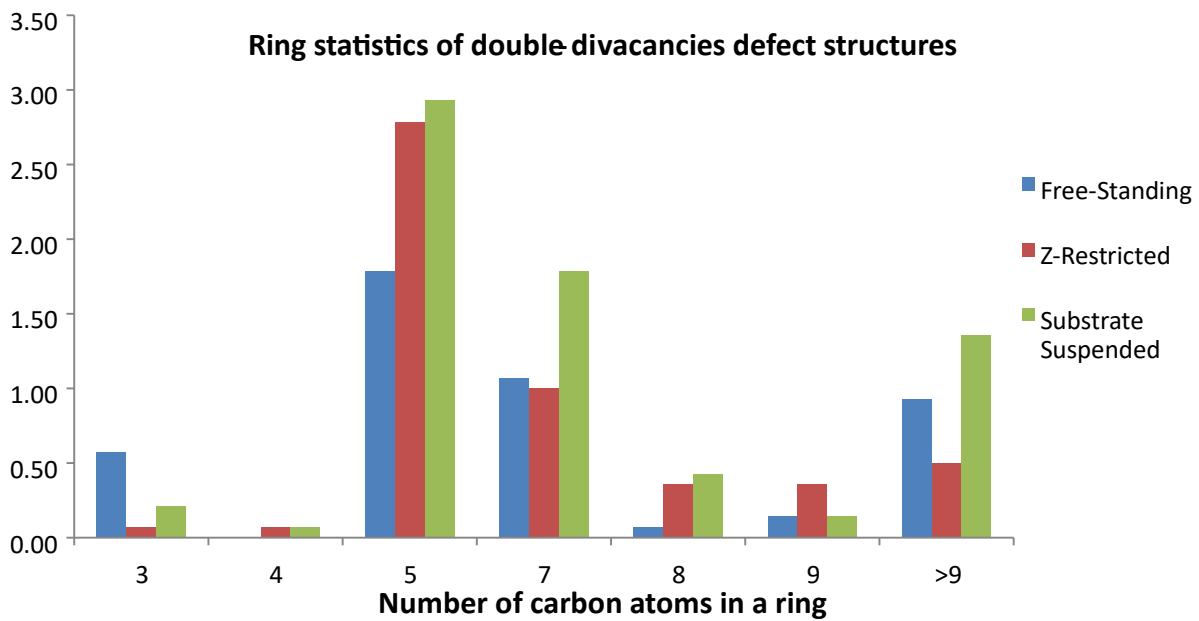

**Figure 5** A histogram of the ring statistics of stable structures formed from graphene with two di-vacancies. The statistics of the cases of free-standing graphene, those with fixed planar configurations and those which are suspended on a rigid planar pristine graphene are represented by the blue, red and green bar, respectively. The y axis represents the number of rings observed for a particular type of N-membered ring averaged over the fourteen distinct initial configurations that were simulated.

**Table 1** A table of the average number of dangling bonds (D.B.) and the average number of non-hexagonal rings that are observed in simulations of defected graphene under three simulation conditions.

|  | Free-standing | Z-restricted | Substrate Suspended |
|---|---|---|---|
| **Number of D.B.** | 4.07 | 1.57 | 4.57 |
| **Number of non-hexagons** | 4.57 | 5.14 | 6.93 |

**Table 2** A table of the standard deviation (Å) of the x, y and z coordinates of a typical atom in a sheet of graphene under the free-standing and substrate suspended conditions. Note that the graphene was placed along the x-y plane at the start of the simulations. The temperature was fixed at 300K.

|  | X/Å | Y/Å | Z/Å |
|---|---|---|---|
| **Free-Standing** | 0.044 | 0.042 | 0.177 |
| **Substrate Suspended** | 0.046 | 0.049 | 0.001 |

Besides having the most number of dangling bonds, substrate suspended defected graphene also produced the largest number of non-hexagonal rings. From FIG. 5, these rings consist of a large proportion of pores, here defined as rings with more than nine members. The porous character and high number of dangling bonds in substrate suspended graphene can be explained by the presence of a periodic spatial potential from the graphene substrate that matches with the lattice structure of the atoms in the defected graphene. Hence, there exist little tendency to move drastically away from the initial configuration, hampering a rearrangement into highly bonded pentagons and heptagons. This explanation is validated by the results of the z-restricted cases, where the only significance difference is the lack of a periodic potential. In those cases, an average of 1.57 dangling bonds was seen as compared to 4.57 dangling bonds in substrate suspended cases. A further reason is that atoms in substrate suspended systems that are adjacent to vacancies tend only to move away from the vacancy centers to the periphery, resulting in an enlargement of the vacancies into pores. In Table 2, the x, y, z motion of free-standing and substrate suspended graphene were compared. The z-motion of the substrate suspended graphene is effectively suppressed and motion is limited to the x, y direction at 300K. This implies that atoms in substrate suspended graphene have less freedom to rearrange themselves via motion in the z direction. This further explains why pores are more prevalent in substrate suspended graphene than free-standing graphene.

Owing to the relatively larger number of dangling bonds, substrate suspended graphene is predicted to be more chemically active than their free-standing counterparts, allowing for more points for reaction. Also, if pore generation via vacancy creation is desired, the graphene should be suspended on a substrate, rather than left free-standing.

For a quantitative comparison between the effectiveness of each simulation condition in producing stable defect structures, ring statistics of the simulations were plotted in FIG. 6. Stable structures are expected to have only a small number of rings that have much more or less than six members. This is because very large or very small rings tend to involve bonds which are too near to each other, making the structure less stable. From FIG. 6, the z-restricted cases are shown to have both the lowest average number of 3 and 4 membered rings and rings with more than 9 members. The z-restricted cases have the second highest average number of 5 and 7 membered rings, which have been deemed to be energetically favored. They also have drastically lower number of dangling bonds (refer to Table 1). These results imply that defects structure formed under z-restriction tend to be more stable than ones formed under the other two conditions.

In view of all these qualitative and quantitative superiority of the z-restricted cases in producing stable defect structures, the z-restricted condition was used to explore stable defects in the simpler systems which are considered in the next two subsections.

## II. Pristine Graphene

Defects in pristine graphene are explored in this subsection. As mentioned in subsection (I), the z-restricted condition was imposed on all simulations. To aid identification of defect structures, the pristine graphene is heated from 0K to a specified temperature in 0.4 ps, then rapidly cooled back to 0K in 0.2 ps, checked for stable defect structures before the procedure was repeated. As before, stable defect structures are deemed to be those with 2 dangling bonds or less. A few trial simulations were carried out to identify the optimal temperature. 2000K was the temperature deemed high enough to allow for a high rate of defect formation while sufficiently low to minimize the creation of huge composite defects that are difficult to characterize and analyze. Due to the z-restriction, the heated atoms only have a coherent oscillation in the x-y plane. The extent of their oscillation increases with an increase in temperature. This trend continues until a defect is formed, then atoms of and around the defect were observed to have larger motion than those further away. The increased motion then triggers a higher rate of defect formation around the first defect. Also, there is an associated high rate of defect transformation. It was realized that this phenomenon contributes significantly to the efficiency of the z-restriction condition in generating different defect structures.

FIG. 6(a) and 6(b) illustrate two possible linear defect structures in pristine graphene. The unit defect in both cases is the 5-7 defect, which usually occur as a doublet 55-77 defect through a Stone-Wales (SW) rotation of a bond. In both cases, an integer multiple of 55-77 are seen, validating the fact that the true defect unit is 55-77. In FIG. 6(a) and 6(b), 55-77 defects are lined up in a linear fashion, along the zigzag and armchair direction, respectively. This implies that stable linearly extended structures are possible in both lattice directions. We will discuss more about the 55-77 defects in subsection (IV).

FIG 6(d) illustrates a possible extension to the linear 55-77 defect structure, which has been proposed as a potential candidate as a metallic wire [38]. It has been formed from a single in-plane rotation by the bond indicated by the green dotted line in FIG. 6(a). The 5555-8-77 structure formed can be further transformed into a defect with alternating 5-8 defects and terminating with a 7 membered ring. Either of the bonds indicated by the red and green dotted lines can undergo SW rotation to add units of 5-8 defects along the zigzag and armchair direction, respectively. As both the linearly extended versions of the 5-8 defect and 55-77 defect have been proposed as parts of a metallic wire structure [38,39], this discovery opens up new possibilities for 2D nano-scale circuitries made with a combination of these defects.

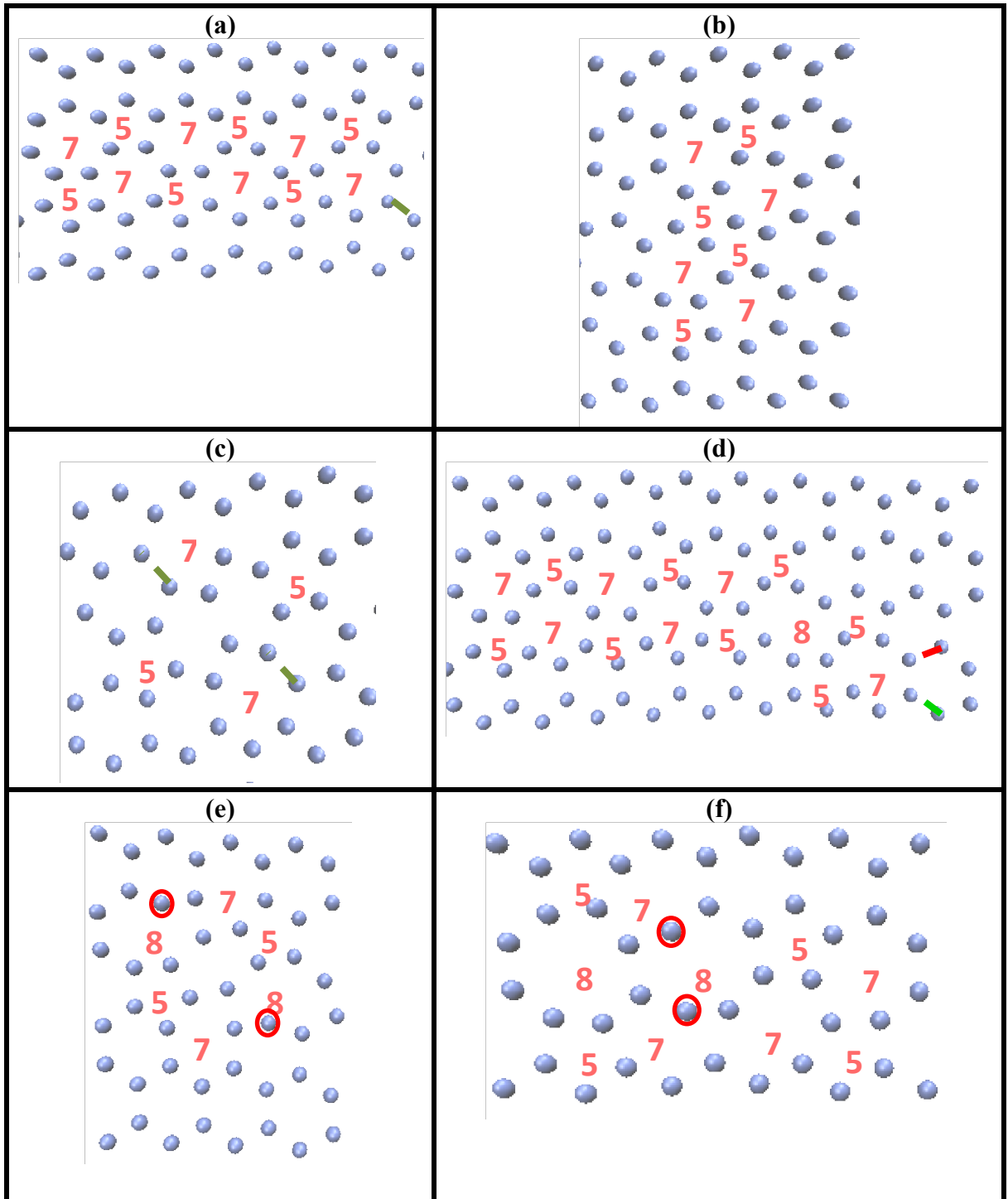

**Figure 6** Structures of defected pristine graphene. **(a)** A line of three 55-77 Stone-Wales defects along the zigzag (ZZ) direction in pristine graphene. **(b)** A line of two 55-77 Stone-Wales defects along the armchair (AC) direction formed in pristine graphene. **(c)** A defect in pristine graphene which consists of two rotated hexagons in the centre with two pentagons and two heptagons positioned symmetrically around it. **(d)** A line of two 55-77 Stone-Wales defects with a 5555-8-77 extension at the right end, in pristine graphene. This structure can be created from three 55-77 defects by rotating the bond indicated by the green dotted line in FIG. 6(a). Further bond rotation can create a line of 5-8 defects along the zigzag (rotating the bond indicated by the red dotted line above) and armchair (rotating the bond indicated by the green dotted line) direction. This implies that we can create an extended line defect consisting of these two types of linear defects by inducing rotation of specific bonds. **(e)** A defect structure

in pristine graphene that consists of two 5-8-7 defects arranged in a ring around a rotated hexagon. This was formed from the structure in FIG. 6(c) by rotating the two bonds indicated by the green dotted lines. Atoms with dangling bonds are indicated by the red circles. 5-8-7 defects have been observed to exist independently. **(f)** A defect structure in pristine graphene. Atoms with dangling bonds are indicated by the red circles.

FIG 6(e) and 6(f) are two defect structures that have two dangling bonds. They are shown for completeness. The defect structure in FIG. 6(c) can be converted to that in 6(e) by SW rotation of the bonds indicated by the green dotted lines. Defect structures with low number of dangling bonds have previously been seen in experiments [36,37]. Moreover, these structures may be stable enough to be present in real graphene for long enough durations to serve as reaction points for functionalization of graphene.

### III.    Mono-vacancy

The same technique as described in subsection (II) has been employed for the cases of graphene with single and double mono-vacancies. The criterion for identification is same as before – structures with two or fewer dangling bonds are deemed stable.

FIG. 7(a) illustrates a defect with a single monovacancy. Although it has three dangling bonds, it is shown because it was the only defect structure observed.

FIG. 7(b-h) illustrate the different defect structures formed from two mono-vacancies. FIG. 7(b-f) are structures with no dangling bonds while each of FIG. 7(g-h) have two dangling bonds.

The defect in FIG. 7(b) was created by a Stone-Wales rotation of the bond labelled by the green dotted line in FIG. 7(f) and the formation of a bond between the two atoms indicated by the red circles. The same bond that had undergone a SW rotation is also labelled in green in FIG. 7(b).

In FIG. 7(c), the common 5-8-5 defect from one divacancy was observed. This implies that the two monovacancies had coalesced to a di-vacancy.

FIG. 7(d) and 7(e) illustrate 5-7 ring structures which have been subject to much experimental investigation and manipulation. Note the difference in lattice orientation between the hexagons in and out of the ring. This illustrates the ring's nature as a grain boundary.

FIG. 7(g) illustrates a healed structure from FIG. 7(f). This structure has the 5-8-7 defect which was also observed in pristine graphene in FIG. 6(e).

FIG 7(h) illustrates a defect structure with two atoms each with a dangling bond. These two atoms can form a bond to create a 55-77 SW defect adjacent to a 5-8-5 defect.

Our simulations produced many experimentally observed structures and some novel structures, which may find applications in novel technologies. The productivity of the simulations validated the z-restriction approach.

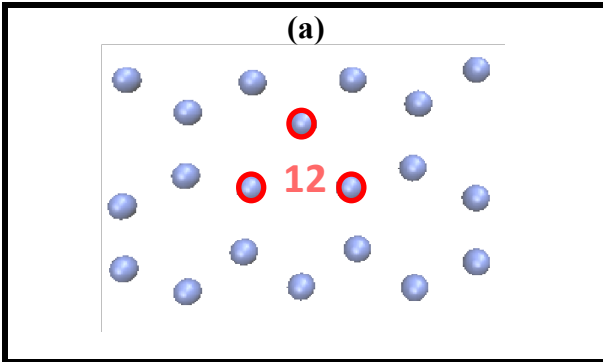
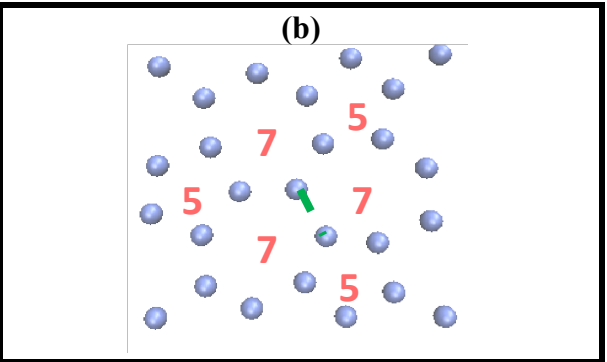
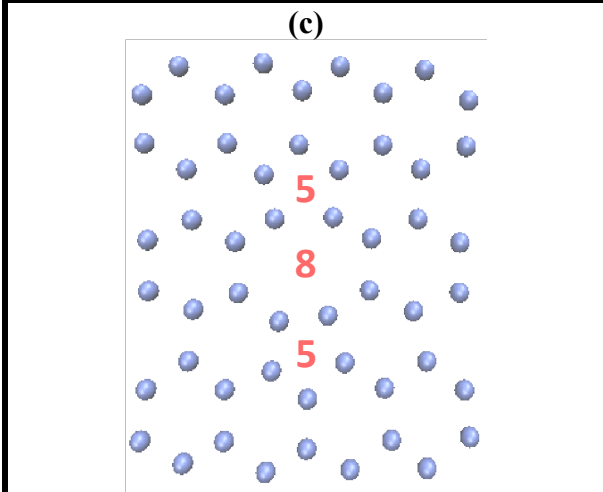
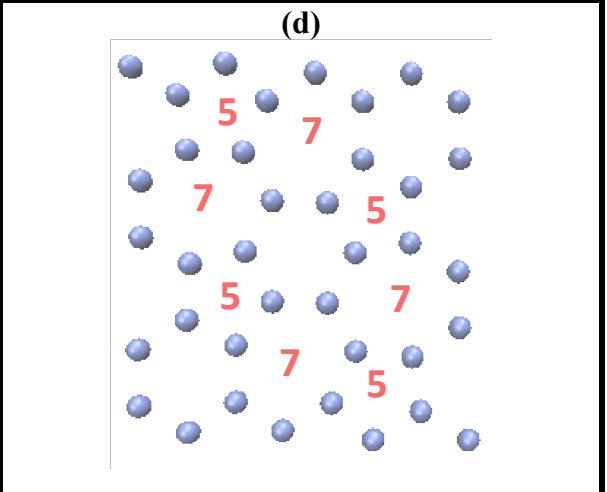
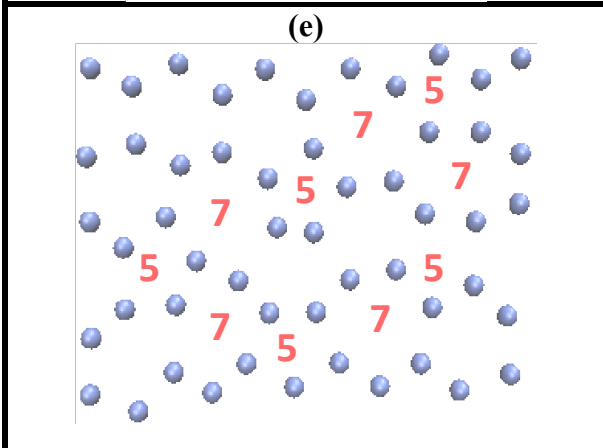
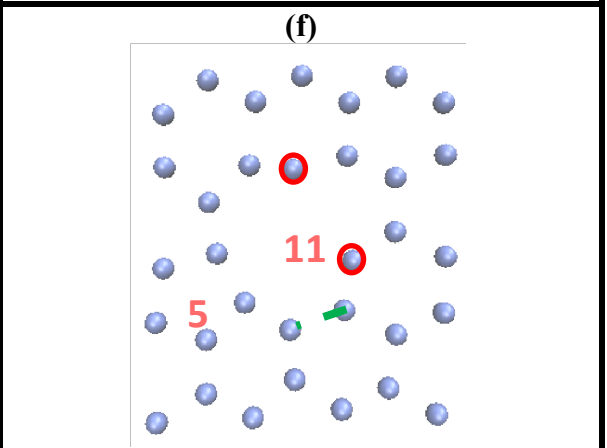
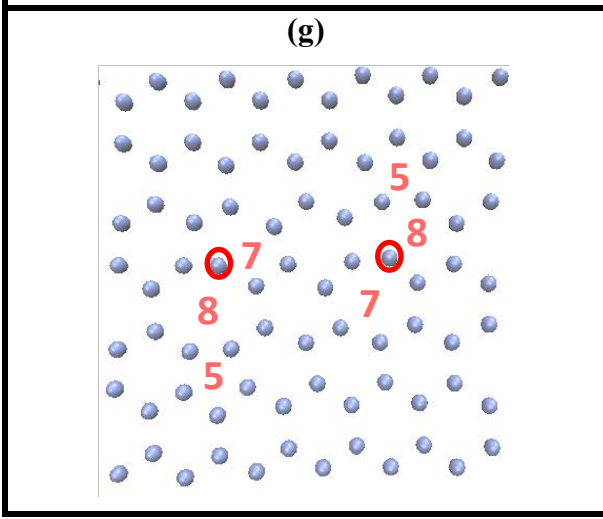
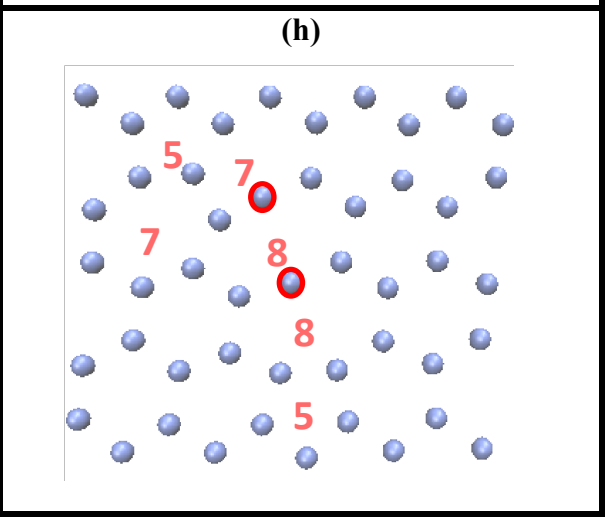

**Figure 7** Structures of graphene with mono-vacancy or Stone-Wales defects. **(a)** A defect formed from a single mono-vacancy and has three dangling bonds. This is the only structure observed formed by a mono-vacancy. Atoms with dangling bonds are indicated by the red circles. **(b)** A 555-777 defect formed from two mono-vacancies. This defect was created by a Stone-Wales rotation of the bond labeled by the green dotted line in FIG. 7(e) and the formation of a bond between the two atoms indicated by the red circles. The bond which had undergone a SW rotation is also labeled in green in this figure. **(c)** A 5-8-5 defect formed from two mono-vacancies. **(d)** A 5555-6-7777 defect formed from two mono-vacancies. This defect consists of 5-7 defects situated around a central hexagon, which has its orientation rotated with respect to external hexagons. **(e)** A 55555-66-77777 defect formed from two mono-vacancies. Note that this structure has one more 5-7 defect and central hexagon than the structure of FIG. 7(c). **(f)** A defect formed from two mono-vacancies and has two dangling bonds. Atoms with dangling bonds are indicated by the red circles. **(g)** Two 5-8-7 defects formed from two mono-vacancies and has two dangling bonds. Atoms with dangling bonds are indicated by the red circles. (h) A defect formed from two mono-vacancies and has two dangling bonds. Atoms with dangling bonds are indicated by the red circles.

## IV.   55-77 Stone-Wales Defects

Further simulations have been done to investigate the properties of extended 55-77 SW defects. Its potential as a nano-scale wire coupled with the seemingly straightforward manner it can be created – through a simple SW rotation by any bond in pristine graphene – made it an interesting target for pristine graphene – made it an interesting target for preliminary studies by simulations. Furthermore, the simulations reported in subsection (II) of pristine graphene suggest that linearly extended 55-77 structures could spontaneously occur under heating in both the zigzag and armchair direction of z-restricted graphene. In contrast, it has been reported that computational simulations of free-standing graphene up to a temperature of 3500K did not produce 55-77 defects [40]. Moreover, the 5-7 edge structure has been reported to be a more energetically stable structure than most other edge structures [41]. Such edge structures have also been seen in experiments [42].

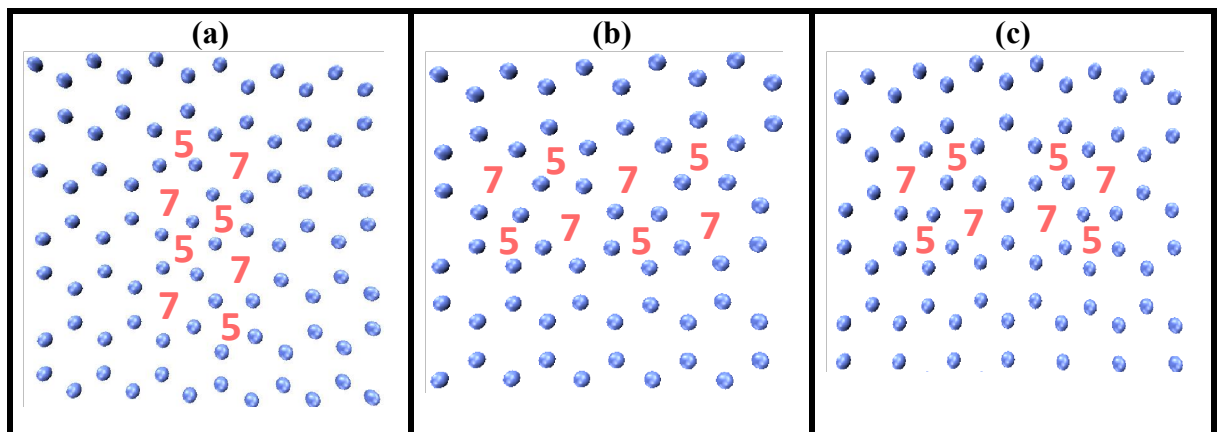

**Figure 8** Graphene structures with extended Stone-Wales defects. **(a)** Two 55-77 defects along the armchair direction, in parallel. **(b)** Two 55-77 defects along the zigzag direction, in parallel. **(c)** Two 55-77 defects along the zigzag direction, in alternating orientation.

In view of the many interesting occurrences of 5-7 defects, linearly extended defects with 55-77 defect as their defect unit were energy minimized to 1 in $10^{10}$ accuracy to extract information about their energetics. As most SW defects are not 2D but involve out-of-plane displacements, the z-restriction was lifted. Bonds in pristine graphene were selectively rotated to generate 55-77 defects at specific locations. The systems investigated involve linearly extended 55-77 defect of different lengths. Previous Density Functional Theory (DFT) calculations [43,44] of the formation energy of a 55-77 defect gave a value of 4.61-5.69 eV, which compares well with our value of 6.12 eV.

FIG. 8(a) and 8(b) show examples of energy minimized structures of two 55-77 defects, placed along different directions. FIG. 8(b) and 8(c) differ by the relative orientation of their defects. Different numbers of 55-77 defects were linearly placed along these directions and orientations and their energy was minimized to calculate their formation energy as a function of length. The result is shown in FIG. 9.

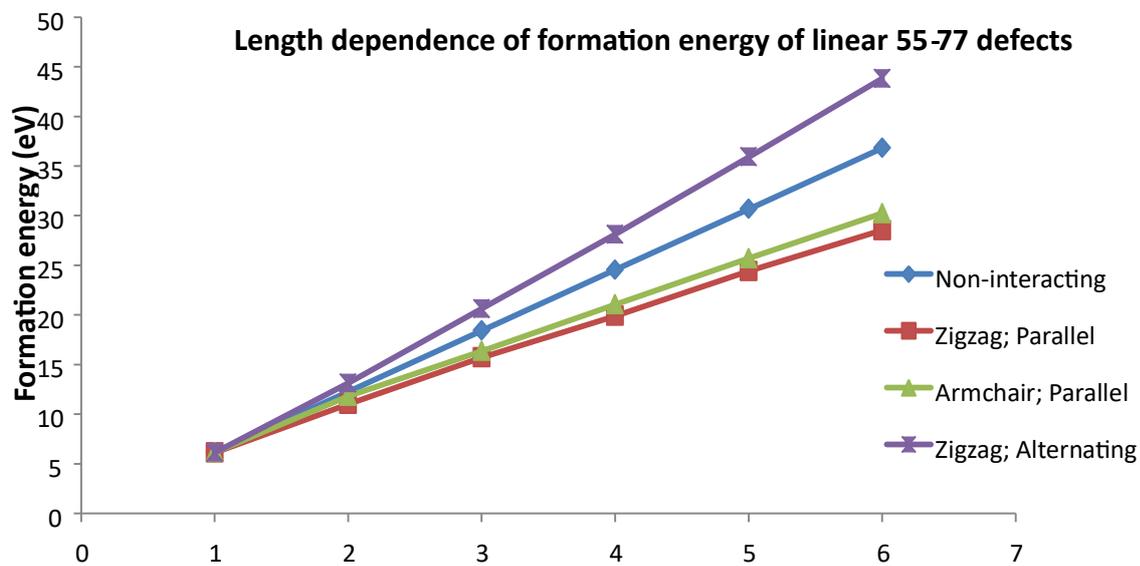

**Figure 9** The length dependence of formation energy of linear 55-77 defects along different directions in pristine graphene and with different relative orientations.

All formation energies in FIG. 9 follow a monotonically increasing trend. As compared to the non-interacting case, defects placed in parallel along the armchair direction and those placed in parallel along zigzag direction are more stable, while those placed in alternating orientations are the least stable. This means that it is energetically more favourable to agglomerate defects in parallel than to keep them apart. Defects placed in parallel along the armchair direction have higher energy than when placed in parallel along the zigzag direction.

Simulations were also done to calculate the formation energies of two 55-77 defects placed along non-equivalent axes. FIG. 10 shows its dependence on separation distance counted in number of hexagons. Defects placed in parallel along the zigzag direction have monotonically increasing energies. This implies that it is most energetically favourable to have two parallel 55-77 next to each other along the ZZ direction. Those placed along the AC direction have an oscillatory behaviour which correlates the fact that they are switching between parallel and non-parallel orientations with increasing separation distance. There is always an increase in energy going from parallel to non-parallel orientation, vice versa. Defects placed in non-parallel orientation along the ZZ direction have decreasing formation energy with increasing distance. This means that such defects are most stable far apart from each other.

However, at a separation distance of 2 and 5 hexagons, there are spikes in the formation energy. The nature of origin of these spikes is unknown and is a topic for future study. There is a convergence in the formation energies at large separation, which confirms the idea that interaction should decrease with an increase in separation.

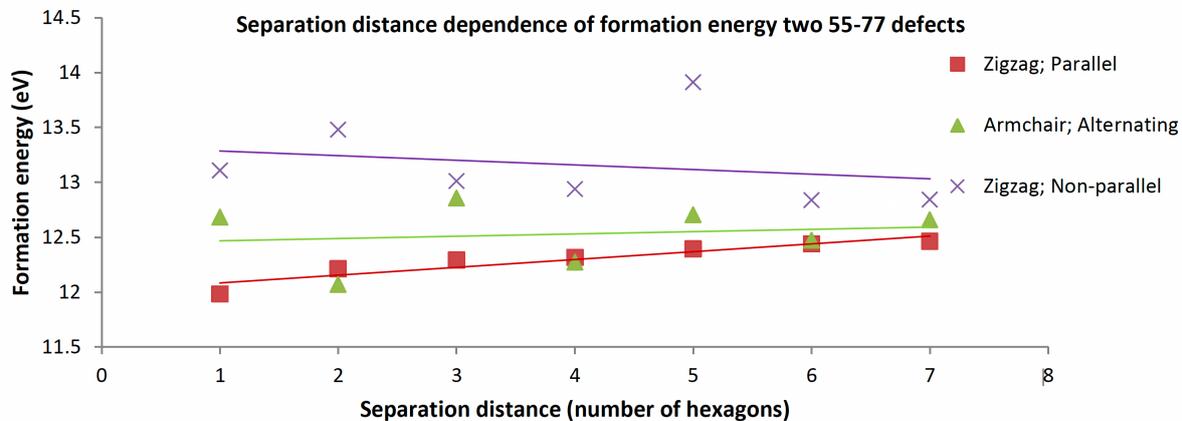

**Figure 10** The separation distance dependence of formation energy of two 55-77 defects. Linear best fit lines are drawn in the same colour as the datapoints they represent.

**Conclusions**

In this work we discussed a molecular dynamics procedure, named DR-MD, to perform an extensive search for stable defect structures of pristine graphene and graphene with one or two mono-vacancies. An exploration of the more complicated case of two divacancies was conducted as test cases to elucidate the best parameters for DR-MD. We report the observation of many structures that have been proposed in theory and/or have been seen in experiments. We also report a few novel defect structures that have yet been reported in literature.

We also explored linear structures created from 55-77 defects. The energetics of the systems were explored and analyzed. To aid our understanding of the dynamics of 55-77 defects, formation energies of two such defects at different separation distance, placed along different directions and with relative orientations were calculated. Further DFT calculations should be conducted to check these values.

These results add to the sum of knowledge on defects in graphene, especially their role in the formation of nanopores. Moreover, it is hoped that the further application and adaptation of DR-MD will prove to be a useful tool in our search for interesting defect structures that may have fascinating properties for novel applications.

**Acknowledgement**

J.W.Y. is grateful to Professors Nicholas Harrison and Matthew Foulkes for the helpful discussions. J.W.Y. is funded by the National Science Scholarship, Agency for Science, Technology and Research (A*STAR), Singapore.

**Author Contributions**

The manuscript was written through contributions of all authors. All authors have given approval to the final version of the manuscript.